\documentclass[manuscript]{aastex}

\usepackage{epsfig}

\shorttitle{IPN Supplement}
\shortauthors{Hurley et al.}

\begin{document}

\title{The Interplanetary Network Supplement to the HETE-2 Gamma-Ray Burst Catalog}

\author{K. Hurley}
\affil{Space Sciences Laboratory, University of California,
7 Gauss Way, Berkeley, CA 94720-7450, U.S.A.}
\email{khurley@ssl.berkeley.edu}

\author{J.-L. Atteia, C. Barraud, A. P\'{e}langeon}
\affil{Laboratoire d'Astrophysique, Observatoire Midi-Pyr\'{e}r\'{e}es,
14 avenue E. Belin, 31400 Toulouse, France}

\author{M. Bo\"{e}r}
\affil{Observatoire de Haute-Provence, 04870 Saint Michel l'Observatoire, France}

\author{R. Vanderspek, G. Ricker}
\affil{Kavli Institute for Astrophysics and Space Research, Massachusetts Institute of Technology,
70 Vassar Street, Cambridge, MA 02139, U.S.A.}

\author{E. Mazets, S. Golenetskii, D. D. Frederiks, V. D. Pal'shin, R. L. Aptekar}
\affil{Ioffe Physico-Technical Institute of the Russian
Academy of Sciences, St. Petersburg, 194021, Russian Federation}

\author{D.M. Smith}
\affil{Physics Department and Santa Cruz Institute for Particle Physics,
University of California, Santa Cruz, Santa Cruz, CA 95064, U.S.A.}

\author{C. Wigger, W. Hajdas}
\affil{Paul Scherrer Institute, 5232 Villigen PSI, Switzerland}

\author{A. Rau, A. von Kienlin}
\affil{Max-Planck-Institut f\"{u}r extraterrestrische Physik,
Giessenbachstrasse, Postfach 1312, Garching, 85748 Germany}

\author{I. G. Mitrofanov, D. V. Golovin, A. S. Kozyrev, M. L. Litvak, A. B. Sanin}
\affil{Space Research Institute, 84/32, Profsoyuznaya, Moscow 117997, Russian Federation}

\author{W. Boynton, C. Fellows, K. Harshman}
\affil{University of Arizona, Department of Planetary Sciences, Tucson, Arizona 85721, U.S.A.}

\author{S. Barthelmy, T. Cline, J. Cummings\altaffilmark{1}, N. Gehrels, H. A. Krimm\altaffilmark{2}} 
\affil{NASA Goddard Space Flight Center, Code 661, Greenbelt, MD 20771, U.S.A.}
\altaffiltext{1}{Joint Center for Astrophysics, University of Maryland, Baltimore County, 1000 Hilltop Circle, Baltimore, MD 21250}
\altaffiltext{2}{Universities Space Research Association, 10211 Wincopin Circle, Suite 500, Columbia, MD 21044}

\author{K. Yamaoka}
\affil{Department of Physics and Mathematics, Aoyama Gakuin University, 5-10-1 Fuchinobe, Sagamihara, Kanagawa 229-8558, Japan}

\author{M. Ohno}
\affil{Institute of Space and Astronautical Science (ISAS/JAXA), 3-1-1 Yoshinodai, Sagamihara, Kanagawa 229-8510, Japan}

\author{Y. Fukazawa, Y. Hanabata}
\affil{Department of Physics, Hiroshima University, 1-3-1 Kagamiyama, Higashi-Hiroshima, Hiroshima 739-8526, Japan}

\author{T. Takahashi}
\affil{Institute of Space and Astronautical Science (ISAS/JAXA), 3-1-1 Yoshinodai, Sagamihara, Kanagawa 229-8510, Japan}

\author{M. Tashiro, Y. Terada}
\affil{Department of Physics, Saitama University, 255 Shimo-Okubo, Sakura-ku, Saitama-shi, Saitama 338-8570, Japan}

\author{T. Murakami}
\affil{Department of Physics, Kanazawa University, Kadoma-cho, Kanazawa, Ishikawa 920-1192, Japan}

\author{K. Makishima\altaffilmark{3}}
\affil{Department of Physics, University of Tokyo, 7-3-1 Hongo, Bunkyo-ku, Tokyo 113-0033, Japan}
\altaffiltext{3}{Makishima Cosmic Radiation Laboratory, The Institute of Physical and Chemical Research (RIKEN),
2-1 Hirosawa, Wako, Saitama 351-0198, Japan}

\author{C. Guidorzi, F. Frontera\altaffilmark{4}, C. E. Montanari\altaffilmark{5}, F. Rossi}
\affil{University of Ferrara, Physics Department, Via Saragat, 1, 44100 Ferrara, Italy}
\altaffiltext{4}{INAF/Istituto di Astrofisica Spaziale e Fisica Cosmica di Bologna, via Gobetti 101, 40129 Bologna, Italy}
\altaffiltext{5}{Istituto IS Calvi, Finale Emilia (MO), Italy}

\author{J. Trombka, T. McClanahan, R. Starr}
\affil{NASA Goddard Space Flight Center, Greenbelt, MD 20771, U.S.A.}

\author{J. Goldsten, R. Gold}
\affil{Applied Physics Laboratory, Johns Hopkins University, Laurel, MD 20723, U.S.A.}

\begin{abstract}
Between 2000 November and 2006 May, one or more spacecraft of the interplanetary
network (IPN) detected 226 cosmic gamma-ray bursts that were also detected
by the FREGATE experiment aboard the HETE-II spacecraft.  During this period, the IPN consisted of up to nine spacecraft,
and using triangulation, the localizations of 157 bursts were obtained.
We present the IPN localization data on these events.
\end{abstract}


\section{Introduction}

The Wide Field X-Ray Monitor (WXM) and Soft X-Ray Camera (SXC) aboard the High Energy Transient Experiment (HETE-2)
mission localized 79 gamma-ray bursts (GRBs) rapidly and precisely between 2001 and 2006 (Vanderspek et al. 2009).  About 1400
more GRBs, however, occurred outside their fields of view and were not detected or localized by them.
In some cases these events were detected by the HETE-2 French Gamma-Ray Telescope (FREGATE) at angles up
to 180 degrees from the detector axis and
identified by onboard and/or ground-based software.  These detections were used
to initiate searches through the data of the spacecraft comprising the interplanetary
network (IPN), and in many cases precise, delayed localizations could be obtained by
triangulation, and multiwavelength counterpart searches were initiated.  Between
2000 November and 2006 May, when these detections occurred, the IPN contained between 4 and 9 spacecraft.  They were,
in addition to HETE:
\it Ulysses, \rm in heliocentric orbit at distances between 670 and 3000
light-seconds from Earth (Hurley et al. 1992); \it Konus-Wind \rm, in various orbits up to around
4 light-seconds from Earth (Aptekar et al. 1995); \it BeppoSAX \rm , in low Earth orbit (Frontera et al. 1997, Hurley et al. 2000); the \it Near-Earth 
Asteroid Rendezvous \rm mission (NEAR), 
in orbit around the asteroid Eros at distances between 775 and 1060 light-seconds
from Earth (Trombka et al. 1999); \it Mars Odyssey\rm, launched in 2001 April and in orbit around Mars starting in
2001 October, up to 1250 light-seconds from Earth (Hurley et al. 2006);
\rm the \it Ramaty High Energy Solar Spectroscopic Imager \rm (RHESSI) in low Earth orbit (Smith et al. 2002); \rm the \it International Gamma-Ray
Laboratory \rm (INTEGRAL), in an eccentric Earth orbit at up to 0.5 light-seconds from Earth (Rau et al. 2005); \rm the \it Mercury Surface, Space Environment, Geochemistry, and Ranging
\rm mission (MESSENGER), launched in 2004 August, but commencing full operation only
in 2007 (Gold et al. 2001); and \it Swift \rm(Gehrels et al. 2004) and \it Suzaku \rm (Takahashi et al. 2007; Yamaoka et al. 2009), both in low Earth orbit. Their timelines are presented in
figure 1.  In this paper, we present the localization data obtained by the IPN for these bursts.

At least two other spacecraft recorded GRB detections during this period, although they were not used
for triangulation and therefore were not part of the IPN.  The \it Rossi X-Ray Timing Explorer \rm (RXTE)  All Sky Monitor
detected and localized some HETE bursts (Smith et al. 1999).  It operated in the low energy X-ray range,
where the light curves of gamma-ray bursts differ significantly from the high energy range where the other IPN instruments
operate, and it was not utilized for triangulation.  The \it Defense Meteorological Satellite Program \rm (DMSP) spacecraft
detected, but did not localize bursts (Terrell et al. 1996, 1998; Terrell and Klebesadel 2004).

\section{Observations}

Whenever a gamma-ray burst or a rapid transient event was detected by
the HETE onboard or ground-based software, a search was initiated in
the data of the IPN spacecraft.  For the spacecraft within a few light-seconds
of Earth, the search window was centered on the HETE trigger time, and its
duration was somewhat greater than the HETE event duration.
For the spacecraft at interplanetary distances, the search window was 
twice the light-travel time to the spacecraft if the event arrival direction
was unknown, which was the case for most events.  If the arrival direction
was known, even coarsely, the search window was defined by calculating the expected arrival time
at the spacecraft, and searching in a window around it.  In addition to these searches, which were initiated
by HETE events, the HETE data were searched whenever an IPN spacecraft detected an event.
Of the more than 1400 events detected by up to 7 IPN spacecraft, 226
were detected by HETE; these are listed in table 1.  
Table 2 shows the number of events observed by each spacecraft in the IPN,
and table 3 gives the number of bursts that were detected by a total
of N spacecraft, where N is between 2 and 9.

\section{Localizations}

When a GRB arrives at two spacecraft with a delay $\rm \delta$T, it may be
localized to an annulus whose half-angle $\rm \theta$ with respect to the
vector joining the two spacecraft is given by 
\begin{equation}
cos \theta=\frac{c \delta T}{D}
\end{equation}
where c is the speed of light and D is the distance between the two
spacecraft.  (This assumes that the burst is a plane wave, i.e. that its
distance is much greater than D.)  The annulus width d$\rm \theta$, is
\begin{equation}
d \rm \theta =c \rm \sigma(\delta T)/Dsin\rm \theta
\end{equation}
where
$\rm \sigma(\delta$T) is the uncertainty in the time delay.  
$\rm \sigma(\delta$T) is generally of the order of 100 ms or more, when
both statistical and systematic uncertainties are considered; thus triangulation
between two near-Earth spacecraft, for which D is at most $\sim$40 ms, does
not constrain the burst arrival direction.  When D is of the order of
several light-seconds (e.g., the distance between \it Konus-Wind\rm \, and
a near-Earth spacecraft), annuli with widths of several degrees can be
obtained; when D is several hundred light-seconds (i.e. an interplanetary
spacecraft and a near-Earth spacecraft), annulus widths of the order of
arcminutes or less are possible.  When two interplanetary spacecraft and a near-Earth
spacecraft observe a GRB, a small error box can be obtained.  Table 4 gives
the number of events observed by 0, 1, and 2 interplanetary spacecraft.

In some cases, no localizations can be obtained which constrain the burst
arrival direction significantly, even though the spacecraft separations are
several light-seconds.  This is due to the fact that one or more of the
spacecraft recorded the event with low time resolution.  

Note that the Swift BAT observes numerous bursts outside its coded field of
view (as indicated by the footnote in table 1).  The Swift data on these events is nevertheless useful for triangulations. 

157 bursts could be localized by the method above; table 5 gives the localization information.  Triangulation annuli are
given in the 8 IPN columns: these are the right ascension and declination
of the annulus center $\alpha, \delta$, the annulus radius R, and the uncertainty in the
radius $\delta$R.  One or two annuli are specified.  In addition to triangulation
annuli, several other types of localization information
are included in this catalog.  The 3 SAX columns give the right ascension, declination,
and radius of the BeppoSAX gamma-ray burst monitor (GRBM) error circle (Frontera et al. 2009). 
The 3 HETE columns give the right ascension, declination, and radius of the WXM or
SXC error circle, whichever is smaller (Vanderspek et al. 2009).  Combining these
error circles with the IPN annuli often results in smaller error regions.  
In one case, GRB040810, the HETE attitude could not be determined precisely,
and consequently the IPN annulus does not intersect the error circle.
We have omitted the HETE error circle coordinates for this burst.
The two Ecliptic columns give the ecliptic latitudes of the bursts, measured
northward (positive) from the ecliptic plane towards the north ecliptic pole.  These are derived by comparing the count
rates of the two \it Konus-Wind\rm \, detectors (Aptekar et al. 1995), and can be considered
to be at the 90--95\% confidence level.  Planet-blocking
is specified by the right ascension and declination of the planet's center and
its radius, in the 3 Planet columns.  When a spacecraft in low Earth or Mars
orbit observes a burst, the planet blocks up to $\approx$ 3.7 sr of the sky.
This is often useful for deciding which of two annulus intersections
is the correct one, or for eliminating portions of a single annulus.
Finally, the Other column gives the right ascension, declination, and
radius of any other localization region, which is obtained in one of
several ways.  For example, error circles can be derived from the intersection
of an IPN annulus and a HETE WXM one-dimensional localization.  Or they
may be from the \it Swift \rm spacecraft
(\url{http://swift.gsfc.nasa.gov/docs/swift/archive/grb\_table/}).  Or they
may be derived from planet-blocking by a second spacecraft in
addition to the data in the Planet column. In this case the
error circle given is the complement of the planet-blocking circle, that
is, a circle whose RA is the RA of the planet plus 180 degrees, whose
declination is the negative of the planet's declination, and whose radius
is 180 degrees minus the planet's angular radius.    
The units of all entries in table 5 are degrees, and all coordinates are J2000.  
For some events, no triangulation was possible, but coarse constraints on
the burst arrival direction can
be derived from planet-blocking, ecliptic latitudes, or both.  This information
is not given here, but information
on these events, as well as the ones in this catalog, may be found at the
IPN website: \url{ssl.berkeley.edu/ipn3/index.html }. 
Figures 2 and
3 show examples of IPN localizations.  Table 6 gives the localization areas in square 
degrees for the bursts in table 5.

\section{Conclusions}

This is the tenth in a continuing series of IPN catalogs; the localization
data for all of them can be found in electronic form at the IPN website.
The IPN is, in effect, a full-time, all-sky monitor, when the duty cycles
and viewing constraints of all its instruments are considered.  Its threshold
for 50\% detection efficiency is about $\rm 6 \times 10^{-7} erg \, cm^{-2}\, or \,
1 \, photon \, cm^{-2}\, s^{-1}$.  Over the HETE-2 mission, 226 bursts were detected by
HETE-FREGATE and at least one other IPN instrument and 157 of them could be localized
to some extent by triangulation.  The more precise and/or rapid localizations were announced via 55 GCN Circulars, 
resulting in multiwavelength counterpart searches.
Regardless of precision and speed of the localizations, however, burst data such as this are useful for numerous studies,
such as searching for indications of activity from previously unknown soft gamma repeaters,
associating supernovae with bursts, or searching for neutrino and gravitational radiation
associated with bursts. 

\section{Acknowledgments}
Between 2001 and 2005, support for the interplanetary network came from the following sources:
JPL Contracts 958056 and 1268385 (Ulysses), MIT Contract SC-R-293291 and
NASA NAG5-11451 (HETE), NASA NNX07AH52G (Konus), NASA NAG5-13080 (RHESSI), NASA NAG5-12614 and  NNG04GM50G (INTEGRAL), 
NASA NAG5-11451 and JPL Contract 1282043 (Odyssey), NASA NNG05GF72G (Swift), NASA NAG5-9126 (BeppoSAX), 
NASA NNX06AI36G (Suzaku), NASA NAG5-9503 (NEAR), and NASA NNX07AR71G (MESSENGER).  In Russia,
this work was supported by the Federal Space Agency of Russia and RFBR grant 09-02-00166a.
C.G., F.F., and E.M. acknowledge financial support by ASI-INAF contract I/088/06/0.

\begin{figure}{}
\epsscale{1}
\includegraphics[scale=0.5]{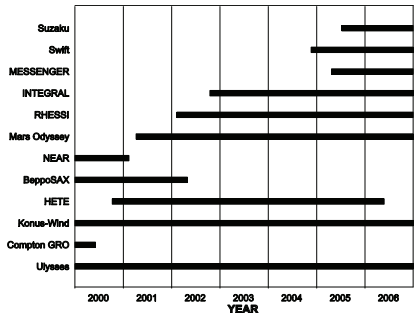}
\caption{The timelines of the missions comprising the interplanetary network
between 2000 and 2006.  During the period when HETE was operational, there were a minimum
3 and a maximum of 8 other missions in the network.  Note that the \it Compton Gamma-Ray Observatory \rm
mission ended before HETE was launched.}\label{fig1}
\end{figure}

\begin{figure}{}
\epsscale{1}
\includegraphics[scale=.50]{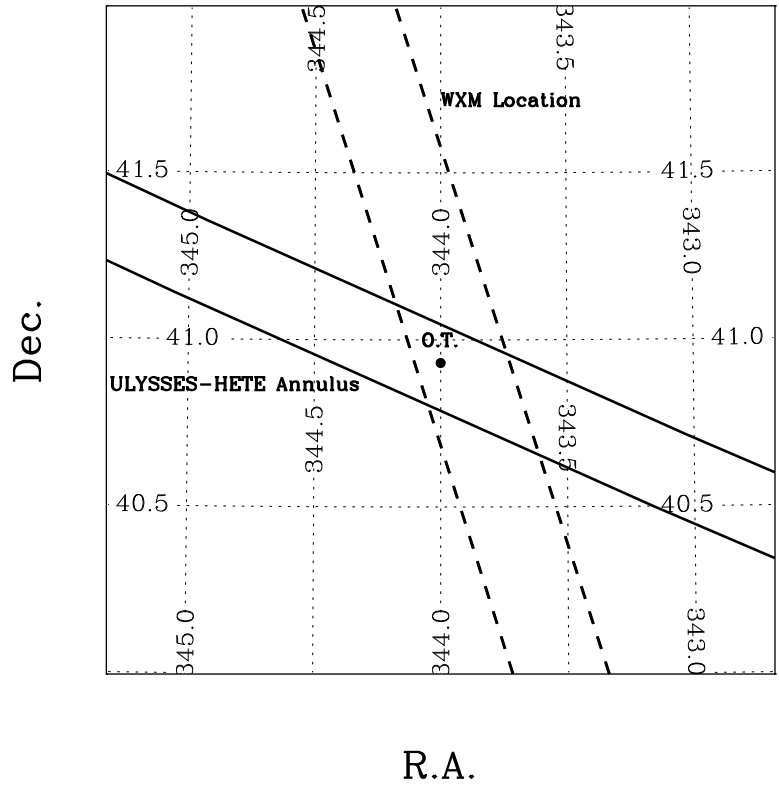}
\caption{Localizations of GRB010921.  The dashed lines show the one-dimensional
WXM localization; the solid lines show the FREGATE/Ulysses annulus.  An optical
transient (OT) was identified by Price et al. (2001).}\label{fig2}
\end{figure}

\begin{figure}{}
\epsscale{1}
\includegraphics[scale=.50]{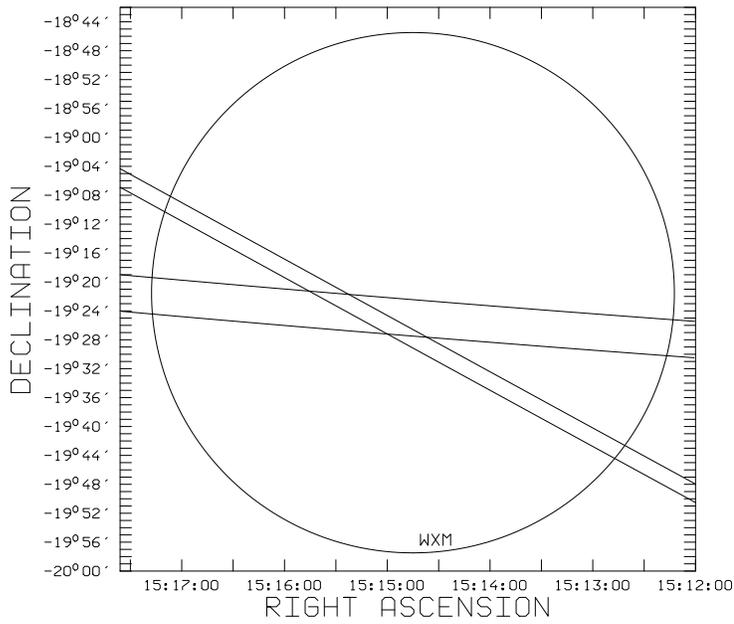}
\caption{Localizations of the short/hard GRB020531.  The wider annulus is
the Ulysses/Odyssey triangulation; the narrower one is the Ulysses/FREGATE
triangulation.  The circle is the
WXM localization.  \it Chandra \rm and optical observations were performed,
but no counterpart was found for this burst.}\label{fig3}
\end{figure}

\appendix

\clearpage

\clearpage

\begin{deluxetable}{cccc}
\tabletypesize{\scriptsize}
\tablecaption{IPN/HETE gamma-ray bursts}
\tablewidth{0pt}
\tablehead{
\colhead{Date} & \colhead{Universal Time}\tablenotemark{a} & \colhead{HETE Identifier} \tablenotemark{b} & 
\colhead{Observed by}\tablenotemark{c} 
}
\startdata
2000 Nov 02 & 15:53:48 & UBID 11414 & Uly,Kon,NEA \\
2000 Nov 05 & 16:25:28 & UBID 11415 & Uly,Kon,NEA \\
2000 Nov 06 & 15:15:15 & UBID 11416 & Uly,Kon,NEA \\
2000 Nov 15 & 12:49:08 & UBID 11418 & Kon \\
2000 Nov 15 & 14:06:41 & UBID 11420 & Kon,SAX \\
2000 Dec 25 & 07:09:20 & UBID 10803 & Uly,Kon,NEA \\
2001 Jan 10 & 19:59:07 & UBID 10809 & Uly,Kon \\
2001 Jan 26 & 09:10:40 & H1487      & Uly,Kon,NEA,XTE \\
\enddata

\tablenotetext{a}{Universal time is the Earth-crossing time of the start of the event}
\tablenotetext{b}{H indicates a triggered burst; UBID stands for
untriggered burst identifier}
\tablenotetext{c}{Kon: \it Konus-Wind\rm; DMS: \it Defense Meteorological Satellite Program\rm;
INT: \it International Gamma-Ray Laboratory\rm;
MO: \it Mars Odyssey\rm; MES: \it Mercury Surface, Space Environment, Geochemistry, and Ranging
\rm mission\rm; NEA: \it Near Earth Asteroid Rendezvous \rm mission; RHE: \it Ramaty High Energy Solar Spectroscopic Imager\rm; 
SAX: \it BeppoSAX\rm;
Suz: \it Suzaku\rm; Swi: \it Swift\rm; Uly: \it Ulysses\rm; XTE: \it Rossi X-Ray Timing Explorer \rm }
\tablenotetext{d}{Burst was outside the coded field of view of the BAT}
\end{deluxetable}

\clearpage

\begin{deluxetable}{cccccccccccc}

\tabletypesize{\scriptsize}
\rotate
\topmargin 1.4in
\tablewidth{8in}
\tablecaption{Number of HETE bursts observed by each spacecraft}
\tablehead{
\colhead{Ulysses} & \colhead{Odyssey} & \colhead{Konus} & \colhead{SAX} & \colhead{HETE} & 
\colhead{NEAR} & \colhead{XTE} & \colhead{RHESSI} & \colhead{INTEGRAL} & \colhead{Swift} & \colhead{MESSENGER} & \colhead{Suzaku} 
}
\startdata
87 & 68 & 181 & 13 & 226  & 5 & 2 & 79 & 69 & 30 & 1 & 14 \\
\tableline
\enddata
\end{deluxetable}

\clearpage

\begin{deluxetable}{ccccccccc}
\tablewidth{0pt}
\tablecaption{Number of HETE bursts observed by N spacecraft}
\tablehead{
\colhead{N=1}           & \colhead{2}      &
\colhead{3}          & \colhead{4}  &
\colhead{5}          & \colhead{6}    &
\colhead{7}   & \colhead{8} & \colhead{9} }
\startdata
0 & 66 & 51 & 66 & 32 & 10 & 1 & 0 & 0 \\
\enddata
\end{deluxetable}

\clearpage

\begin{deluxetable}{ccc}
\tablewidth{0pt}
\tablecaption{Number of bursts observed by N interplanetary spacecraft}
\tablehead{
\colhead{N=0}           & \colhead{1}      &
\colhead{2} }
\startdata
104 & 83 & 39 \\
\enddata
\end{deluxetable}


\begin{deluxetable}{cccc}
\tabletypesize{\scriptsize}
\tablecaption{Error box areas}
\tablewidth{0pt}
\tablehead{
\colhead{Date} & \colhead{Universal Time} & \colhead{Area, square degrees}  
}
\startdata

02 Nov 2000 & 15:53:48 & 1.57E-01 \\
05 Nov 2000 & 16:25:28 & 9.63E-04 \\
06 Nov 2000 & 15:15:15 & 1.09E-03 \\
15 Nov 2000 & 12:49:08 & 3.75E+03 \\
15 Nov 2000 & 14:06:41 & 1.00E+03 \\
25 Dec 2000 & 07:09:20 & 5.09E-02 \\
10 Jan 2001 & 19:59:07 & 4.19E+01 \\
26 Jan 2001 & 09:10:40 & 2.57E-01 \\
\enddata
\end{deluxetable}

\end{document}